\documentclass[11pt]{article}

\usepackage[margin=1in]{geometry}
\usepackage{setspace}
\usepackage{amsmath,amssymb}
\usepackage{url}
\usepackage{graphicx}
\usepackage{tikz}
\usetikzlibrary{arrows.meta,positioning}
\usepackage[numbers,sort&compress]{natbib}
\usepackage{microtype}
\usepackage[hidelinks]{hyperref}
\newcommand{\secref}[1]{\nameref{#1}}

\newcommand{\FPscale}{0.286}
\newcommand{\FPrepsA}{1000}
\newcommand{\FPrepsB}{20}
\newcommand{\FPTB}{2,000}

\newcommand{\FPNBmin}{10^{2}}
\newcommand{\FPNBmax}{10^{5}}

\newcommand{\FPsparseThreshold}{50}
\newcommand{\FPminEpisodes}{15}
\newcommand{\FPminEpisodesN}{10^{2}}
\newcommand{\FPminEpisodesMu}{10^{-5}}
\newcommand{\FPzeroEpisodeRuns}{6}
\newcommand{\FPpanelALoss}{25}
\newcommand{\FPpanelAFreeze}{33}
\newcommand{\FPepAtMostOneThousand}{77}

\newcommand{\FPepMedianHundredThousand}{2}

\newcommand{\FPextMedianThousand}{28}

\newcommand{\FPextMedianHundredThousand}{66}

\newcommand{\FPfreezeDThousand}{73}

\newcommand{\FPfreezeDHundredThousand}{89}
\newcommand{\FPlostPThousand}{74}
\newcommand{\FPlostPHundredThousand}{89}
\newcommand{\FPperiodThousandEminusTwo}{14}
\newcommand{\FPperiodThousandEminusThree}{24}
\newcommand{\FPperiodThousandEminusFour}{92}
\newcommand{\FPperiodThousandEminusFive}{519}

\newcommand{\FPfracHiMin}{19}
\newcommand{\FPfracHiMax}{45}

\title{How Withheld Punishment Sustains Anti-Democratic Behavior\\ in an Evolutionary Game}
\author{Chad M. Topaz\\[3pt]
{\small Williams College, Williamstown, MA, USA}\\
{\small QSIDE Institute, Williamstown, MA, USA}\\
{\small University of Colorado Boulder, Boulder, CO, USA}\\
{\small \texttt{cmt6@williams.edu}}}
\date{\today}

\begin{document}

\maketitle
\raggedbottom
\setcounter{secnumdepth}{0}

\begin{abstract}
\noindent Democratic backsliding is often framed as a contest between defenders and disruptors of democratic institutions. That framing can miss a third response, withholding punishment from disruptors while penalizing defenders who confront them. We model the three responses as strategies in an evolutionary game, track their shares with replicator dynamics, and assume defenders beat disruptors head-to-head but have a payoff disadvantage among non-punishers even when disruptors are absent. If the third response cannot spread among disruptors, a payoff condition draws the population toward the classical May--Leonard heteroclinic cycle, a rock--paper--scissors succession in which disruption repeatedly becomes rare, then widespread. If that response can, a threshold on its payoff advantage over disruptors determines whether all three coexist or defenders alone vanish. In finite-population simulations, sustained resurgence requires reintroducing lost strategies. Our contribution is to show how two separately documented political tendencies, if combined, can sustain disruption despite defenders' head-to-head advantage over disruptors.
\end{abstract}

\noindent\textbf{Keywords:} evolutionary game theory, replicator dynamics, cyclic dominance, heteroclinic cycle, antisocial punishment, democratic backsliding

\bigskip
\section{Introduction}
\label{sec:intro}

In recent decades, elected leaders in countries including Hungary, Poland, Turkey, and Venezuela have eroded democratic institutions from within by packing courts, rewriting electoral rules, and capturing independent media~\citep{Bermeo2016,Scheppele2018}. Such erosion can leave democracy's outward forms intact~\citep{LevitskyZiblatt2018}, and it forms part of a broader wave of autocratization, a shift away from democratic rule that is now usually gradual~\citep{LuhrmannLindberg2019}. This erosion of democratic institutions from within, \emph{democratic backsliding}~\citep{WaldnerLust2018}, is often framed as a contest between actors who defend those institutions and actors who disrupt them~\citep{LevitskyWay2002,Bermeo2016,PiersonSchickler2020}. Disruptors often persist, and sometimes prevail. We ask whether the public's response can keep disruptors in play even when defenders would win a direct, two-way contest.

To see why that response matters, consider a leader who fills a court with loyalists, reducing the court's ability to check the leader's power. The move weakens a democratic constraint, often without breaking any written law, and we call moves of this kind norm-breaking. Part of what the move costs the leader depends on how people react. If they tolerate or excuse it, the leader loses little standing with them. Now suppose the opposition answers with denunciations, boycotts, or street protests. Part of what that costs the opposition depends on the same people. If they see the resulting conflict, rather than the court packing, as the problem, the confrontation costs the opposition support. These public responses can themselves become more or less common, changing the costs of both norm-breaking and opposition.

We study a public that combines these two tendencies, tolerance of norm-breaking and disapproval of confrontation. Such a public need not oppose democracy. Its response may reflect moderation, partisan loyalty, conflict aversion, exhaustion, or indifference. We represent defending institutions, disrupting them, and this combined response as strategies in an evolutionary game. Officials and citizens share one population, and anyone in it can switch among the three strategies. We use the labels \emph{defenders}, \emph{disruptors}, and \emph{non-punishing public} to indicate which strategy an actor currently uses. The last label refers only to actors using the third strategy, not to the public as a whole.

We now examine this third and least familiar strategy. It pairs weak punishment for norm-breaking with a penalty on visible confrontation. We take the evidence for each in turn and then state the assumption that links them.

The evidence for weak punishment comes from survey experiments. In experiments in which respondents choose between hypothetical candidates for office, a candidate who adopts an undemocratic position loses an estimated 12 percentage points of vote share overall. In the scenarios most like real elections, in which the candidates differ in party and their policies align with their partisanship, the estimated loss falls to about 3.5 points. Voters also punish the other side far more readily than their own~\citep{GrahamSvolik2020}. Voters are not indifferent to every violation. They do penalize several specific violations across party lines, though partisan division is sharpest over voter-identification laws~\citep{CareyEtAl2022}. A separate experiment varied which party held power in a hypothetical scenario. Respondents were more supportive of norm-eroding policies when their own party held power, a pattern its authors call democratic hypocrisy. This partisan difference more than doubled among respondents randomly prompted to list ways in which the opposing party threatened them, and it was larger among those who identified strongly with their own party as a social group~\citep{SimonovitsMcCoyLittvay2022}. These studies show that punishment for norm-breaking can be weak and partisan, although voters penalize some violations across party lines.

The evidence for a penalty on visible confrontation comes from studies of incivility and protest. Uncivil televised debate lowers political trust after even brief exposure, most of all among conflict-averse viewers~\citep{MutzReeves2005}. Uncivil argument shown in television close-up lowers the perceived legitimacy of the opposing side~\citep{Mutz2007}. Beyond television, extreme protest tactics, those seen as harmful to others or highly disruptive, reduce public support for a movement and its cause~\citep{FeinbergWiller2020}, and in the United States of the 1960s, protester-initiated violence produced electoral backlash while nonviolent protest did not~\citep{Wasow2020}. These studies show costs for some confrontational tactics, particularly incivility and violence. The television studies show costs to political trust and to the perceived legitimacy of opponents; the protest studies show costs to the confronting side's own support. They measure reactions to confrontation in general, though, not to the confronting of norm-breakers in particular. The model assumes that a penalty of this kind carries over to confrontation in defense of democratic norms. These findings do not establish a general penalty for nonviolent protest or for institutional defense through criticism, oversight, and litigation.

The evidence just reviewed supports the two tendencies separately. But the third strategy in our model requires a further behavioral assumption: that the two tendencies go together, so that the actors who spare norm-breakers are, to a first approximation, the same actors who penalize those who confront them. The assumption has not been tested directly, and we return to it in \secref{sec:discussion}.

Several motives could produce one or both tendencies. One possibility is conflict aversion, under which actors who dislike confrontation might overlook a breach and resent those who confront it. Conflict-averse citizens take part less in confrontational forms of participation such as protest, campaigning, and contentious discussion, though their turnout is not significantly lower~\citep{UlbigFunk1999}, and exposure to differing political views within one's own network is associated with lower participation~\citep{Mutz2002AJPS}. Many people avoid settings they expect to be combative, and a 2018 U.S. survey found an exhausted majority weary of partisan conflict~\citep{GroenendykKrupnikovRyanConnors2025,MoreInCommon2018}, a distaste that is partly for the conflict itself rather than for the opposing party~\citep{KlarKrupnikovRyan2018}.

A second possibility is partisan loyalty, under which actors excuse a co-partisan's norm-breaking and could also resent those who make an issue of it. Partisanship and affective polarization, the warmth people feel toward their own side alongside hostility toward the other~\citep{IyengarEtAl2019,Mason2016}, can dull the impulse to punish a co-partisan's anti-democratic conduct~\citep{GrahamSvolik2020,SimonovitsMcCoyLittvay2022}. Perceived threat can likewise raise tolerance for illiberal measures~\citep{FeldmanStenner1997,HetheringtonSuhay2011}, though it bears only on the tolerance of norm-breaking. The model will not specify which motive produces the behavior; it takes the behavior itself as given.

We now step back to the model as a whole, with all three strategies in play. To describe how the share of actors using each strategy changes over time, we assign each strategy a payoff, a numerical measure of its performance against the current mix of strategies. Payoffs can reflect electoral support, approval, or access to office. Defenders have a payoff advantage when disruptors make up nearly all of the population and a disadvantage when the non-punishing public does. We assume that this disadvantage holds whether or not disruptors are present. A standing cost of defense that non-punishers avoid can produce it, whereas a penalty paid only when a defender confronts a disruptor cannot. Our results depend on this assumption, and \secref{sec:methods} discusses its basis. Disruptors, in turn, have a payoff advantage when the non-punishing public makes up nearly all of the population, but not when defenders do. We use replicator dynamics, equations that describe how payoff differences change strategy shares. A strategy already in use becomes more common when its payoff exceeds the population's average payoff and less common when its payoff falls below that average~\citep{TaylorJonker1978,HofbauerSigmund2003}. These equations can arise when actors copy others whose strategies are performing better.

Our analysis uses classical results on replicator dynamics. Three-strategy systems with cyclic dominance, in which each strategy beats one of the others and loses to the third, have been studied for decades. The resurgence pattern we analyze is the May--Leonard heteroclinic cycle~\citep{MayLeonard1975,Hofbauer1994}. The role of a third strategy also has precedents in evolutionary models of cooperation. In public goods games, players can contribute to a shared project or benefit from it without contributing. Allowing players to abstain from the project can replace the dominance of free riders over contributors with a cycle involving contributors, free riders, and abstainers~\citep{HauertEtAl2002a,HauertEtAl2002b,SemmannEtAl2003}. Cycling can persist when the full set of punishment strategies is available, including punishment of contributors by abstainers~\citep{RandNowak2011}. Our contribution is to show how two separately documented political tendencies, if combined, can sustain disruption despite defenders' head-to-head advantage over disruptors.

Our results describe two ways in which the combined response can sustain disruption. In the first, the \emph{exploitation regime}, the pairwise advantages form a cycle. Under a classical condition on the payoffs, the population approaches the May--Leonard cycle and disruptors repeatedly resurge. Defenders beat back disruptors. With disruptors scarce, defenders still bear the standing costs of defense, and the non-punishing public, which avoids those costs and penalizes confrontation, grows at their expense. With defenders scarce, disruptors exploit the public's withheld punishment and spread again. With each round, the population comes closer to states in which everyone uses one strategy, and it stays near those states longer.

This prediction assumes a continuum of actors, in which a behavior can become arbitrarily rare without disappearing. In a finite population, the last actor using a behavior can switch to another strategy. We therefore simulate populations with and without reintroduction of lost strategies. The simulations show how strategy loss interrupts resurgence and measure how often disruptor majorities recur when lost strategies can return.

The \emph{accommodation regime} provides a second way for the combined response to sustain disruption. There, the non-punishing strategy can spread when rare among disruptors, and disruptors can likewise spread when rare among non-punishers. The outcome depends on the public strategy's payoff advantage over disruptors when it is rare among them. At or above a threshold in that advantage, the two settle into a stable coalition that excludes defenders; below it, all three behaviors coexist. Here, \emph{coalition} means stable evolutionary coexistence of the two strategies, not a modeled political alliance. At the threshold, the equilibrium with all three behaviors meets the equilibrium without defenders and the two exchange stability, a transcritical bifurcation. The threshold lies strictly below the defenders' head-to-head advantage, so defenders can be excluded even when their advantage over disruptors exceeds the public's advantage from accommodation. More generally, once a non-punishing third strategy is present, an enforcer's pairwise advantage over the behavior it is meant to suppress no longer determines the global outcome.

Four families of models lie close to ours. Two model politics directly, and two provide evolutionary precedents for our strategies and dynamics. The first family is the game theory of political conflict. Game-theoretic accounts of backsliding model the strategic choices of an aspiring autocrat and of the actors who might constrain them, such as voters, courts, and opposition parties, and they ask whether institutional constraints survive those choices~\citep{GrilloEtAl2024}. Two dynamical models are closer to ours in form, an evolutionary clash-of-cultures game in which modest, fair, and greedy claims compete for a contested resource~\citep{ArceSandler2003}, and a model of radicalization and deradicalization through indoctrination and attacks~\citep{ShortEtAl2017}. None of these models includes the response we study, withholding punishment from norm-breakers while penalizing those who confront them.

The second family is the dynamics of opinions and networks. Opinion-dynamics models represent each individual's position on an issue axis and revise it under social influence. These models can produce consensus, clusters of opinions, or polarization, depending on who interacts and how those interactions change opinions~\citep{FriedkinJohnsen1999,Deffuant2000,HegselmannKrause2002,FlacheMacy2011,FlacheEtAl2017}. Recent work in this journal studies related questions at the level of networks, including affective polarization as a structural property of online interaction networks~\citep{LermanEtAl2024} and the coevolution of attitudes and network ties under politically motivated influence~\citep{LipatovEtAl2025}. It also examines the structure of decision groups that lets even a polarized population make coherent collective decisions~\citep{HebertDufresneEtAl2025}. These models describe ideological positions and network structure, and our question concerns behavioral responses. A position on an issue axis does not by itself say how a person responds to norm-breaking, whether by confronting it, joining in it, or excusing it while disapproving of those who confront it. We will take these three responses as strategies.

The third family, and the closest formal precedent for our model, is the public goods game with optional participation. In a public goods game, players choose whether to contribute to a shared pool that is multiplied and divided equally among the participating players. Contributing benefits the group, but the benefit a player receives from their own contribution is smaller than its cost, so those who free-ride, the defectors, do better than those who contribute, the cooperators. In the voluntary version of the game, a third strategy, the loner, abstains from the enterprise for a small fixed income, and this converts the pairwise dominance of defectors over cooperators into a rock--paper--scissors cycle. Defectors spread when rare among cooperators, loners spread when rare among defectors, and cooperators replace loners when only those two strategies are present~\citep{HauertEtAl2002a,HauertEtAl2002b}. The cycling has been observed in the laboratory, where staged starting conditions produce each predicted succession~\citep{SemmannEtAl2003}. The last step of the cycle differs from a strict invasion. A participant who finds no co-players receives the loner's income, so at the all-loner state a rare cooperator's proportional growth rate is zero rather than positive. Cooperators still increase from any positive share in a population of cooperators and loners, but motion near the all-loner state is much slower than near the other two single-strategy states~\citep{HauertEtAl2002b}. The exploitation regime of our model will instead have, at every single-strategy state, a strictly positive growth rate for the strategy that invades and a strictly negative one for the strategy that is driven out.

The loner result shows that adding a third strategy can keep the winner of a two-way contest from taking over. Our main structural claim, that the non-punishing public keeps disruptors in play despite the defenders' head-to-head advantage, is an instance of the same mechanism. The public is not simply a loner, though. A loner only abstains, and in the loner game cooperators beat loners in a population of those two strategies alone. Our public abstains from enforcement and also penalizes the enforcers, so it beats defenders in a population of those two strategies alone.

The two games also behave differently in the long run. In the loner game, the payoffs come from groups rather than pairs, and its replicator dynamics are Hamiltonian, with a quantity that remains constant along each trajectory, so no state in which all three strategies are present attracts the population. Unless it starts at equilibrium, a population using all three strategies either cycles indefinitely, with an amplitude fixed by its starting composition, around an equilibrium at which all three coexist, or approaches the all-loner state, depending on the parameters~\citep{HauertEtAl2002b}. Our payoffs will be pairwise and linear, and under the product condition derived in \secref{sec:results} the cycle of the exploitation regime will be an attracting May--Leonard heteroclinic cycle.

The fourth family is the evolutionary theory of punishment. Three ideas connect this literature to our public strategy. The first is costly enforcement. Defenders can be read as norm enforcers, and people do enforce norms at their own expense. In experiments on third-party punishment, many observers were willing to pay to sanction violations that did not affect their own payoffs~\citep{FehrFischbacher2004}. Models of \emph{altruistic punishment}, punishment that costs the punisher and benefits the group, ask how such enforcement can arise and persist~\citep{BoydEtAl2003}.

The second idea is the \emph{second-order free rider}, who benefits from enforcement without paying for it and so undermines the enforcers~\citep{PanchanathanBoyd2004,Fowler2005}. Allowing players to abstain helps costly punishment gain a foothold~\citep{Fowler2005,HauertEtAl2007}, whereas in a finite population with rare mutation, compulsory participation favors defectors~\citep{HauertEtAl2007}.

The third idea is \emph{antisocial punishment}, the sanctioning of those who contribute. Our public resembles a second-order free rider who additionally imposes a cost on the punishers, which is analogous to antisocial punishment. In public goods experiments across sixteen participant pools, antisocial punishment was widespread. In some pools it was strong enough to cancel the benefit of punishment to cooperation, and it was harsher where norms of civic cooperation and the rule of law were weaker~\citep{HerrmannEtAl2008}. Retaliation can also undermine enforcement. When punished players can retaliate, cooperators become less willing to punish free riders, and cooperation breaks down~\citep{Nikiforakis2008}. In evolutionary models, admitting the full set of punishment strategies, rather than only punishment aimed at defectors, can remove the benefit of punishment to cooperation and favor substantial antisocial punishment~\citep{RandEtAl2010,RandNowak2011}.

The closest behavioral precedent for our non-punishing public strategy is a loner who punishes contributors. In the optional public goods game with the full set of punishment strategies, most antisocial punishers are loners, who abstain from the enterprise and then sanction its contributors, and the sanctioning protects them from invasion by cooperators~\citep{RandNowak2011}. That is the response of our non-punishing public, which abstains from defending institutions and penalizes those who do. In this political interpretation, we assume that defenders beat disruptors head-to-head and ask whether that advantage is enough for a stable defender victory. We also analyze the accommodation regime, in which the public and disruptor strategies can each invade the other, a relation with no counterpart in the basic loner game.

The rest of this paper is organized as follows. \secref{sec:results} states the model, with three strategies and two payoff regimes, analyzes the two regimes, and examines strategy loss and reintroduction in finite populations. \secref{sec:discussion} interprets the two regimes, discusses evidence relevant to distinguishing them, and states the model's scope and limitations. \secref{sec:methods} describes the actors and strategies and the political motivation of the payoff assumptions, reviews replicator dynamics as a model of how behaviors spread through imitation and verifies the properties of the dynamics that \secref{sec:results} uses, explains when cyclic competition produces sustained resurgence and why finite populations can interrupt that pattern, and describes the finite-population simulations.

\section{Results}
\label{sec:results}

\subsection{Strategies, payoffs, and dynamics}
\label{sec:model}

We now specify the model, which describes how the three behavioral responses become more or less common. We first name the strategies and represent the population by the share of actors using each, then introduce payoffs and replicator dynamics, and then state the payoff assumptions. These assumptions define the exploitation and accommodation regimes, which differ in whether the non-punishing response can spread when rare among disruptors. \secref{sec:methods} describes the actors and strategies in more detail, verifies the properties of the dynamics used here, explains how imitation can produce these dynamics, and gives the political motivation for each payoff assumption.

The three strategies are defending democratic institutions, disrupting them through norm-breaking, and withholding sanction from norm-breakers while penalizing those who visibly confront them. We label them $D$, $A$, and $P$, and we call the actors using them defenders, disruptors, and the non-punishing public. All three strategies are available to actors in a single population that pools institutional roles, as \secref{sec:methods} explains. Any actor can switch strategies.

Let $x_D$, $x_P$, and $x_A$ denote the shares of actors using the defender, public, and disruptor strategies at time $t$, and let $x(t)=(x_D,x_P,x_A)$ collect them. The vector $x(t)$ lies on the simplex, the set of possible population compositions,
\begin{equation}
\Delta^2 = \{x\in\mathbb{R}^3_{\ge0}: x_D + x_P + x_A = 1\}.
\end{equation}
Its three vertices $e_D=(1,0,0)$, $e_P=(0,1,0)$, $e_A=(0,0,1)$ are the states in which the whole population plays a single strategy. Each edge is a segment on which one strategy is absent. The edge $\{x_A=0\}$, for instance, represents populations with no disruptors, and away from its endpoints both defenders and the non-punishing public are present. Because the shares sum to one, any two of them determine the third, so the simplex is a triangle and the dynamics on it are planar, describable by two variables. Its relative interior, where every $x_i>0$, is the triangle with its boundary, the edges and vertices, removed. As the shares change, the state traces a trajectory, and a state that does not change is an equilibrium. We next specify how the population shares change within this triangle.

In the fixed payoff matrix $M \in \mathbb{R}^{3\times 3}$, the entry $M_{ij}$ gives the payoff to an actor using strategy $i$ in an interaction with someone using strategy $j$. A payoff represents benefits that help a strategy spread, such as electoral support, approval, or access to office or resources. We express all payoffs on a common scale, in units of a proportional growth rate. They do not measure social welfare or democratic value.

We assume well-mixed interactions, in which each actor samples a partner uniformly from the population, without regard to institutional role. The probability that this partner uses strategy $j$ is $x_j$. Strategy $i$ therefore has expected payoff $\pi_i(x)=\sum_j M_{ij}x_j=(Mx)_i$, and averaging over all actors gives $\bar{\pi}(x)=\sum_i x_i\pi_i(x)=x^\top M x$. The population shares follow replicator dynamics~\citep{TaylorJonker1978,HofbauerSigmund2003},
\begin{equation}
\dot{x}_i = x_i\big[(Mx)_i - x^\top M x\big], \qquad i \in \{D,P,A\}.
\label{eq:replicator}
\end{equation}
Here $\dot{x}_i$ is the rate of change of strategy $i$'s share. For $x_i>0$, its proportional growth rate $\dot{x}_i/x_i$ equals its payoff minus the population average. Its share increases when this difference is positive.

Under \eqref{eq:replicator}, the shares continue to sum to one, an initially absent strategy remains absent, and a positive share stays positive at every finite time, although it may approach zero as time increases. \secref{sec:methods} verifies these properties. The state therefore remains on the simplex. Every vertex is an equilibrium, and each edge is forward-invariant, meaning that a trajectory that starts on an edge stays there.

Adding a constant to every entry in a column of the payoff matrix leaves the replicator dynamics unchanged~\citep{HofbauerSigmund2003}, as \secref{sec:methods} also verifies, so we can set the diagonal of the payoff matrix to zero. Starting from an unnormalized matrix $U$, we subtract $U_{jj}$ from every entry in column $j$ to obtain
\begin{equation}
M_{ij}=U_{ij}-U_{jj}.
\label{eq:U}
\end{equation}
The resulting matrix has zero diagonal. The six off-diagonal entries, one for each direction on each edge of the simplex, then fix the entire model. Each normalized entry compares the payoff of strategy $i$ against strategy $j$ with what $j$ earns against itself. As the population approaches the vertex $e_j$, where everyone uses strategy $j$, the proportional growth rate of a rare strategy $i$ approaches $M_{ij}$. This limiting rate is its invasion fitness, and $i$ can invade when $M_{ij}>0$ and cannot when $M_{ij}<0$.

A negative $M_{ij}$ does not tell us whether an actor using $i$ earns less against $j$ than against another actor using $i$. That comparison would require $U_{ij}<U_{ii}$, which the normalized matrix does not determine. We therefore use the words \emph{harm}, \emph{gain}, \emph{exploitation}, and \emph{accommodation} throughout in this relative, evolutionary sense, as statements about invasion rather than about welfare. Welfare comparisons of this kind motivate some of our payoff assumptions, but the results do not require those comparisons to hold.

We specify the payoff signs edge by edge. Two edges have the same signs in both regimes; the regimes differ on the public--disruptor edge. We write positive entries as $b$ and negative entries as $-a$, with $a,b>0$ and with the subscript naming the strategy that receives the payoff. We write $X\succ Y$ when $X$ can invade $Y$ but $Y$ cannot invade $X$. On their shared edge, $X$ then prevails whenever both strategies are initially present.

On the defender--disruptor edge, we assume that defenders, in isolation, roll back disruptors, so $M_{DA}=b_D>0$ and $M_{AD}=-a_A<0$, that is, $D \succ A$. This assumption is deliberately favorable to the defenders, because it lets us ask whether head-to-head dominance is enough for a stable defender victory.

On the public--defender edge, we assume $M_{DP}=-a_D<0$ and $M_{PD}=b_P>0$, that is, $P\succ D$. The two entries are separate comparisons, which \secref{sec:methods} spells out. The positive entry alone makes the all-defender state unstable. The negative entry makes the defender strategy decline when rare among non-punishers, which the cycle of the exploitation regime requires and which enters the threshold of the accommodation regime through $a_D$. On this edge, no disruptor is present, so neither comparison involves the confrontation of a disruptor.

On the public--disruptor edge, we assume in both regimes that disruptors can invade a population of non-punishers, so $M_{AP}>0$. The regimes differ in whether the non-punishing strategy can invade a population of disruptors in turn.

In the \emph{exploitation} regime, $M_{PA}=-a_P<0$, so the non-punishing response cannot invade a population of disruptors, and with $M_{AP}=b_A>0$ we have $A\succ P$. In terms of the unnormalized payoffs, the assumption is $U_{PA}<U_{AA}$. \secref{sec:methods} gives the welfare comparison that motivates the regime's name.

In the \emph{accommodation} regime, the non-punishing strategy can invade disruptors as well, so the two are mutually invadable. For this regime, we write the two positive entries as $M_{PA}=c_P$ and $M_{AP}=c_A$. The accommodation regime assumes mutual invasibility, $U_{PA}>U_{AA}$ and $U_{AP}>U_{PP}$, without requiring an absolute gain to either party.

With rows and columns ordered $(D,P,A)$, the payoff matrix of the exploitation regime is
\begin{equation}
M^{(1)}=
\begin{pmatrix}
0      & -a_D &  b_D \\
b_P    &  0   & -a_P \\
-a_A   &  b_A &  0
\end{pmatrix},
\qquad a_D,a_P,a_A,b_D,b_P,b_A>0,
\label{eq:M1}
\end{equation}
and the accommodation regime replaces the $P$--$A$ block to give
\begin{equation}
M^{(2)}=
\begin{pmatrix}
0      & -a_D &  b_D \\
b_P    &  0   &  c_P \\
-a_A   &  c_A &  0
\end{pmatrix},
\qquad a_D,a_A,b_D,b_P,c_P,c_A>0.
\label{eq:M2}
\end{equation}

Figure~\ref{fig:schematic} shows the two competitive structures. In the exploitation regime, the three relations form strict cyclic dominance, a cycle with a clear winner on every edge. In the accommodation regime, the $P$--$A$ edge becomes mutually invadable while $P\succ D$ and $D\succ A$ remain. The signs fix these structures, and the remaining subsections of \secref{sec:results} show how the magnitudes decide the long-run outcomes within each.

\subsection{The exploitation regime: a resurgence cycle}
\label{sec:exploitation}
Pairwise cyclic dominance does not by itself determine what happens when all three strategies are present. We first establish the direction of change on each edge, then determine whether populations containing all three strategies approach an equilibrium, approach the boundary cycle, or oscillate with a fixed amplitude. Figure~\ref{fig:regimes} illustrates the attracting-equilibrium and attracting-cycle cases.

Under $M^{(1)}$, expected payoffs are
\begin{equation}
\pi_D(x)=-a_D x_P + b_D x_A, \qquad
\pi_P(x)= b_P x_D - a_P x_A, \qquad
\pi_A(x)= -a_A x_D + b_A x_P.
\label{eq:payoffs1}
\end{equation}
On an edge where only strategies $i$ and $j$ are present, with shares $u$ and $1-u$, the population average is $u\pi_i+(1-u)\pi_j$, and \eqref{eq:replicator} reduces to $\dot u = u(1-u)(\pi_i-\pi_j)$. Between the vertices, the strategy with the higher payoff increases in share. If the payoff difference keeps the same strict sign throughout the edge, the favored strategy's share approaches one whenever both strategies are initially present, and we say that it eliminates the other. On the $D$--$P$ edge ($x_A=0$), $\pi_P-\pi_D=b_Px_D+a_D(1-x_D)>0$. On the $P$--$A$ edge ($x_D=0$), $\pi_A-\pi_P=b_Ax_P+a_P(1-x_P)>0$. On the $D$--$A$ edge ($x_P=0$), $\pi_D-\pi_A=b_D(1-x_D)+a_Ax_D>0$. Each difference is positive throughout its edge, so no edge has an equilibrium between its vertices. The boundary flow therefore forms the directed cycle
\begin{equation}
D \longrightarrow P \longrightarrow A \longrightarrow D.
\label{eq:cycle}
\end{equation}
The arrows show the direction of population change along each edge, so that $D\to P$ means that $P$ eliminates $D$. This is the cyclic dominance assumed above. The three single-strategy equilibria and the boundary trajectories connecting them form a heteroclinic cycle, and the question is whether trajectories from the interior approach it. In political terms, the non-punishing public displaces confrontational defenders, disruptors displace the non-punishing public, and defenders displace disruptors, each in a population of those two strategies alone.

An interior equilibrium is a point at which all three strategies earn the same payoff. Because the payoffs in \eqref{eq:payoffs1} are linear in $x$, the conditions $\pi_D=\pi_P=\pi_A$ and $x_D+x_P+x_A=1$ form a linear system, which is nonsingular and has the unique solution
\begin{equation}
x_D^\ast=\frac{a_Pa_D+a_Pb_A+b_Ab_D}{S},\quad
x_P^\ast=\frac{a_Aa_P+a_Ab_D+b_Pb_D}{S},\quad
x_A^\ast=\frac{a_Aa_D+a_Db_P+b_Ab_P}{S},
\label{eq:interioreq}
\end{equation}
where $S>0$ is the common denominator, equal to the sum of the three numerators, so that the shares sum to one. Each numerator is a sum of products of the positive parameters, so all three shares are positive and $x^\ast$ lies strictly inside the simplex.

To find whether nearby trajectories approach $x^\ast$ or move away from it, we eliminate $x_D=1-x_P-x_A$, which reduces the interior dynamics to a planar system in the coordinates $(x_P,x_A)$. We then compute the Jacobian $J$ of that system at $x^\ast$, the matrix of first derivatives of the two rates of change, which gives a linear approximation to the motion near the equilibrium. Small departures from $x^\ast$ decay when both eigenvalues of $J$ have negative real part and grow when both have positive real part. The trace and determinant of $J$ are
\begin{equation}
\operatorname{tr}J(x^\ast)=\frac{a_Aa_Pa_D-b_Ab_Pb_D}{S},
\qquad
\det J(x^\ast)=S\,x_D^\ast x_P^\ast x_A^\ast>0.
\label{eq:trace}
\end{equation}
The eigenvalues, and hence the trace and determinant, do not depend on which coordinate is eliminated. Because $\det J(x^\ast)>0$, both eigenvalues have real parts with the sign of the trace whenever the trace is nonzero, so $x^\ast$ repels exactly when $\operatorname{tr}J(x^\ast)>0$, that is, when
\begin{equation}
a_Pa_Aa_D>b_Pb_Ab_D.
\label{eq:product}
\end{equation}
The eigenvalues are complex for every choice of positive parameters. To verify this, multiply the three numerators in \eqref{eq:interioreq}. Their product is $S^2\det J$ and exceeds $(a_Da_Pa_A)^2+(b_Db_Pb_A)^2$. This sum exceeds $(a_Da_Pa_A-b_Db_Pb_A)^2=S^2(\operatorname{tr}J)^2$, so $(\operatorname{tr}J)^2<4\det J$. Under either strict inequality, nearby trajectories therefore spiral, outward under \eqref{eq:product} and inward under the reverse inequality. At equality in \eqref{eq:product}, the eigenvalues are purely imaginary, and the linearization alone does not decide the behavior near $x^\ast$; we return to this case below.

The same inequality governs the boundary. It is equivalent to $\rho_D\rho_P\rho_A>1$, where $\rho_i$ is the ratio at the vertex $e_i$ of the contraction rate to the expansion rate. At each vertex, one rare strategy grows and the other declines. Their limiting proportional growth rates are the two off-diagonal entries in the corresponding column of $M$. These invasion fitnesses are also the Jacobian's eigenvalues in the two edge directions. The positive rate is the expansion rate along the outgoing edge of the cycle, and the magnitude of the negative rate is the contraction rate along the incoming edge. Explicitly,
\begin{equation}
\rho_D=\frac{-M_{AD}}{M_{PD}}=\frac{a_A}{b_P},\qquad
\rho_P=\frac{-M_{DP}}{M_{AP}}=\frac{a_D}{b_A},\qquad
\rho_A=\frac{-M_{PA}}{M_{DA}}=\frac{a_P}{b_D}.
\label{eq:rho}
\end{equation}
For example, near the all-defender vertex, rare disruptors decline at proportional rate $a_A$, while the public strategy, when rare, grows at proportional rate $b_P$. Thus $\rho_D=a_A/b_P$. The product condition $\rho_D\rho_P\rho_A>1$ is the standard criterion for a heteroclinic cycle to attract~\citep{Hofbauer1994}, and in this game the cycle neither attracts nor repels at equality. Inequality~\eqref{eq:product} therefore does two jobs at once. It makes the interior equilibrium repel, and it makes the boundary cycle attract.

To move from local stability to long-run behavior, we use one structural result. A change of variables to the ratios $x_P/x_D$ and $x_A/x_D$, with a positive rescaling of time, turns the three-strategy replicator equation into a planar Lotka--Volterra system, the classical two-species model of ecology. Planar Lotka--Volterra systems admit no isolated periodic orbits, also called limit cycles. Their periodic orbits occur only as a neutral family surrounding a center, an equilibrium enclosed by a continuum of closed trajectories~\citep{Hofbauer1981,Zeeman1980,Bomze1983}. The interior equilibrium of our system is such a center only at equality in \eqref{eq:product}, where the trace vanishes. Under either strict inequality, the system has no periodic orbits at all. The argument does not extend to four or more strategies, where isolated periodic orbits can occur.

Under \eqref{eq:product}, the interior equilibrium repels and the boundary vertices are saddles, each with one attracting and one repelling direction, because their two edge eigenvalues, the entries of $M$ appearing in \eqref{eq:rho}, differ in sign. Each edge runs from one vertex to the next with no equilibrium in between, so the boundary is exactly the cycle $D\to P\to A\to D$. Back in the simplex, which is compact and forward-invariant, consider the set of points that a non-equilibrium interior trajectory approaches as $t\to\infty$, its $\omega$-limit set. By the Poincar\'e--Bendixson theorem, it is an equilibrium, a periodic orbit, or a closed chain of equilibria joined by connecting trajectories. The limit set cannot be $x^\ast$, because that equilibrium repels. Nor can it be a vertex, since only trajectories on the corresponding incoming boundary edge converge there. Periodic orbits have already been ruled out, leaving the boundary cycle, the closed chain of vertices and edges, as the only possibility~\citep{Zeeman1980,Bomze1983,Hofbauer1981}.

Because the cycle attracts, a non-equilibrium interior trajectory spends ever longer periods near each vertex. Write $T_D$, $T_P$, and $T_A$ for the times spent near $e_D$, $e_P$, and $e_A$ on one circuit, and $T_D'$ for the time spent near $e_D$ on the next. As the trajectory approaches the cycle, the ratios of successive residence times converge, with $T_P/T_D\to a_A/b_A$, $T_A/T_P\to a_D/b_D$, and $T_D'/T_A\to a_P/b_P$. Each limit compares the contraction rate at one vertex with the expansion rate at the next. Multiplying the three limits gives $\rho_D\rho_P\rho_A>1$. Thus, on successive circuits, the time spent near any given vertex asymptotically increases by this factor, and all three residence times grow without bound. The ratio $T_A/T_P\to a_D/b_D$ shows that disruptor-dominated episodes last longer relative to the preceding public-dominated episodes when $b_D$ is smaller or $a_D$ is larger. The population therefore returns to periods in which almost everyone uses the disruptor strategy, and these periods grow longer.

For a trajectory approaching the cycle, even the time average of the population composition fails to converge. This average, $\frac{1}{T}\int_0^T x(t)\,dt$, records the strategy shares averaged over $[0,T]$. Each new period of dominance lasts long enough to shift it substantially. Its accumulation points, the limits obtained along different sequences of ever later times $T$, form the boundary of a triangle inside the simplex whose vertices are set by the expansion and contraction rates~\citep{Gaunersdorfer1992}. We summarize:

\medskip
\noindent\textbf{Proposition 1 (Resurgence).} \emph{Under $M^{(1)}$ with $a_P a_A a_D > b_P b_A b_D$, the unique interior equilibrium is repelling and every non-equilibrium interior trajectory approaches the boundary heteroclinic cycle $D\to P\to A\to D$. Residence times near the vertices grow without bound, and the time average of the state does not converge but accumulates on the boundary of a triangle~\citep{Gaunersdorfer1992}.}

\medskip
Proposition~1 assembles the classical analyses of May--Leonard-type heteroclinic cycles in competition and replicator systems~\citep{MayLeonard1975,Hofbauer1994,Gaunersdorfer1992}. Our contribution is the political reading. The non-punishing public converts a contest that defenders would win head-to-head into a cycle. Under \eqref{eq:product}, non-equilibrium interior trajectories approach this cycle. Disruptors repeatedly become dominant, and their periods of dominance grow longer (Figure~\ref{fig:regimes}a); in this deterministic model, the resurgence continues indefinitely.

Under the reverse strict inequality $a_Pa_Aa_D<b_Pb_Ab_D$, the cycle repels and the interior equilibrium attracts nearby trajectories. The limit-set argument above, with the boundary cycle now repelling, shows that every interior trajectory converges to that equilibrium. The three behaviors therefore settle into a stable mixture, with disruptors persisting at the share $x_A^\ast$ (Figure~\ref{fig:regimes}b).

At equality, the interior equilibrium is a center surrounded by the neutral family of closed orbits and is not asymptotically attracting; away from $x^\ast$, the three behaviors then oscillate indefinitely with an amplitude fixed by the initial condition. Only the strict inequality in \eqref{eq:product} produces the attracting heteroclinic cycle with lengthening residence times.

\subsection{The accommodation regime: the invasion threshold}
\label{sec:accommodation}
In the accommodation regime, the public and disruptors can coexist without defenders. Our question is whether that defender-free state withstands invasion by defenders or gives way to coexistence of all three strategies. The answer turns on a single threshold. We first determine whether defenders can invade the coalition, and then establish the outcome from arbitrary starting populations that contain all three strategies.

Under $M^{(2)}$, the $D$--$A$ and $D$--$P$ edges behave as before, with $D$ eliminating $A$ and $P$ eliminating $D$, but the $P$--$A$ edge changes. The public strategy has invasion fitness $M_{PA}=c_P>0$ among disruptors, and the disruptor strategy has invasion fitness $M_{AP}=c_A>0$ among the public, so each can invade the other. On that edge ($x_D=0$, $x_A=1-x_P$), the payoffs are $\pi_P=c_Px_A$ and $\pi_A=c_Ax_P$. The dynamics reduce to $\dot x_P = x_P(1-x_P)\big(c_P(1-x_P)-c_A x_P\big)$. This equation has a unique equilibrium between the two vertices,
\begin{equation}
x^\dagger=\Big(0,\;\frac{c_P}{c_P+c_A},\;\frac{c_A}{c_P+c_A}\Big).
\label{eq:coalition}
\end{equation}
Trajectories that start between the two vertices approach $x^\dagger$ from either side. We call $x^\dagger$ the coalition. The word refers to coexistence of the public and disruptors at an equilibrium that is stable along the $P$--$A$ edge, not to a modeled alliance; the model contains no bargaining or coordination. Stability along the $P$--$A$ edge does not establish stability against invasion by defenders.

At the coalition, the defender strategy's proportional growth rate when rare is $\pi_D(x^\dagger)-\bar\pi(x^\dagger)$. Because the two present strategies earn equal payoff at $x^\dagger$, $\bar\pi(x^\dagger)=\pi_P(x^\dagger)=\pi_A(x^\dagger)$. Substituting the coordinates gives $\pi_D(x^\dagger)=(b_Dc_A-a_Dc_P)/(c_P+c_A)$ and $\pi_P(x^\dagger)=c_Pc_A/(c_P+c_A)$, so
\begin{equation}
\lambda_D(x^\dagger)=\frac{H}{c_P+c_A},\qquad H\equiv b_Dc_A-c_P\,(c_A+a_D).
\label{eq:H}
\end{equation}
Defenders can invade the coalition exactly when $H>0$, and rare defenders decline when $H<0$. Written as a threshold on the public strategy's invasion advantage among disruptors, $H<0$ reads
\begin{equation}
c_P>c_P^{\ast}\equiv\frac{b_Dc_A}{c_A+a_D}=\frac{b_D}{1+a_D/c_A},
\label{eq:threshold}
\end{equation}
and $c_P^{\ast}<b_D$ strictly, because $a_D>0$.

The stronger condition $c_P>b_D$ is a useful comparison. Under it, the public strategy outperforms defenders throughout the simplex, because $\pi_P-\pi_D=b_Px_D+a_Dx_P+(c_P-b_D)x_A$ is then positive at every point. Because the shares sum to one, this gap is bounded below by $\min\{b_P,a_D,c_P-b_D\}>0$, so $d\log(x_D/x_P)/dt=\pi_D-\pi_P$ stays negative and bounded away from zero. Along an interior trajectory, the ratio $x_D/x_P$ therefore tends to zero, and defenders disappear. The exact threshold is lower, and the band $c_P^{\ast}\le c_P\le b_D$ between the two is where the exact analysis adds to this simple argument.

The invasion calculation in \eqref{eq:H} is local. It gives the proportional growth rate of the defender strategy when rare near the coalition, not the outcome from other compositions of the population. For populations that initially contain all three strategies, the following accommodation subsections establish defender exclusion at or above $c_P^{\ast}$ and three-strategy coexistence below it. At the threshold itself, the invasion fitness is zero, so a separate nonlinear calculation is needed there.

\subsection{The accommodation regime: equilibria and local stability}

An equilibrium with all three strategies present must satisfy the equal-payoff conditions $\pi_D=\pi_P=\pi_A$, two independent linear equations,
\begin{equation}
b_Px_D+a_Dx_P+(c_P-b_D)x_A=0,\qquad -(a_A+b_P)x_D+c_Ax_P-c_Px_A=0,
\label{eq:equalpay}
\end{equation}
whose common solutions form the line spanned by the cross product of their coefficient vectors,
\begin{equation}
\begin{aligned}
N_D&=b_Dc_A-c_P(c_A+a_D)=H,\\
N_P&=b_D(a_A+b_P)-a_Ac_P,\\
N_A&=a_D(a_A+b_P)+b_Pc_A.
\end{aligned}
\label{eq:N}
\end{equation}
An interior equilibrium is a point of this line scaled so that its components are positive and sum to one, which requires $N_D$, $N_P$, and $N_A$ to share a sign. Since $N_A>0$, this requires $N_D>0$ and $N_P>0$. If $H>0$, then $c_P<c_P^{\ast}<b_D$, so $N_P=a_A(b_D-c_P)+b_Db_P>0$. An interior equilibrium therefore exists exactly when $H>0$, and then
\begin{equation}
x^{\ast}=\frac{(N_D,N_P,N_A)}{S},\qquad S=N_D+N_P+N_A>0.
\label{eq:xstar2}
\end{equation}
An interior equilibrium exists exactly when defenders can invade the coalition. As $c_P$ increases to $c_P^{\ast}$, so that $H$ decreases to zero, the point defined by \eqref{eq:xstar2} moves to the $P$--$A$ edge, where $N_D=0$ and a short calculation gives $N_P/N_A=b_D/(c_A+a_D)=c_P/c_A$. It reaches the coalition there and, as $c_P$ increases further, leaves the simplex. Whether the interior equilibrium attracts nearby trajectories when it exists is the next question.

An equilibrium is a sink when both eigenvalues of its Jacobian have negative real part; nearby trajectories then converge to it. For a planar system, this occurs when the trace is negative and the determinant is positive. For $x^\ast$, the trace follows from an identity that holds for any diagonal-normalized three-strategy game with an interior equilibrium,
\begin{equation}
\operatorname{tr}J(x^{\ast})=-\bar\pi(x^{\ast}),
\label{eq:traceidentity}
\end{equation}
where $J$ is the Jacobian of the dynamics restricted to the simplex, as in the analysis of the exploitation regime. In the present regime, $\bar\pi(x^\ast)=\pi_P(x^\ast)=b_Px^\ast_D+c_Px^\ast_A>0$ whenever $x^\ast$ exists, so
\begin{equation}
\operatorname{tr}J(x^{\ast})=-\frac{a_Aa_Dc_P+b_Db_Pc_A}{S}<0.
\label{eq:trace2}
\end{equation}
The determinant is $\det J(x^{\ast})=S\,x^\ast_Dx^\ast_Px^\ast_A>0$, by the same direct computation that gave \eqref{eq:trace}. The interior equilibrium is therefore a sink whenever it exists. Trajectories approach it along spirals when $(\operatorname{tr}J)^2<4\det J$ and without spiraling otherwise, and both cases occur.

The identity \eqref{eq:traceidentity} has a short proof. Let $J_{\mathrm{full}}$ be the $3\times3$ Jacobian of \eqref{eq:replicator} at $x^\ast$, with entries $(J_{\mathrm{full}})_{ij}=x^\ast_i\big[M_{ij}-\bar\pi-(M^\top x^\ast)_j\big]$. Because the diagonal of $M$ vanishes, its trace is $-2\bar\pi$. The vector $x^\ast$ itself is an eigenvector of $J_{\mathrm{full}}$ with eigenvalue $-\bar\pi$. As a direction, $x^\ast$ changes the sum of the coordinates, so it does not describe motion within the simplex. The other two eigenvalues govern motion within the simplex, and their sum is $-\bar\pi$. In the exploitation regime, the identity recovers \eqref{eq:trace}.

The coalition's stability follows from \eqref{eq:H}. Its two eigenvalues are $-c_Pc_A/(c_P+c_A)$ along the edge and the defender invasion fitness $H/(c_P+c_A)$ transverse to it. For $H<0$, both are negative, and the coalition is a sink, hyperbolic in the sense that no eigenvalue has zero real part. For $H>0$, the coalition is a saddle. Its stable manifold, the set of starting points whose trajectories converge to it, is the $P$--$A$ edge without its vertices, and its unstable direction points into the interior.

At $H=0$, the defender invasion fitness vanishes, so linearization does not determine whether small defender shares grow or decline. Write $d=x_D$ for the defender share and $r=x_P/(1-d)$ for the public's share among non-defenders, so that the coalition is the point $d=0$, $r=c_P/L_0$ with $L_0=c_P+c_A$. Nearby trajectories approach an invariant curve through the coalition, the center manifold, and the slower change in the defender share is governed by the dynamics along this curve. The curve is $r=c_P/L_0+(a_A+b_P)\,d/L_0+O(d^2)$, and on it
\begin{equation}
\dot d=-\frac{S_0}{L_0}\,d^2+O(d^3),
\label{eq:centermanifold}
\end{equation}
where $S_0>0$ is the value of $S$ at $H=0$. Defenders therefore still disappear at the threshold, but their share decreases in proportion to $1/t$ rather than exponentially, with $d\sim L_0/(S_0t)$.

The local picture is now complete. For $H>0$, the interior equilibrium attracts nearby trajectories and the coalition is vulnerable to defenders. For $H\le0$, the coalition attracts nearby interior trajectories, exponentially when $H<0$ and algebraically at $H=0$. What remains is to rule out other long-run outcomes for trajectories that start elsewhere in the simplex.

\subsection{The accommodation regime: global convergence}

By the Poincar\'e--Bendixson theorem, the $\omega$-limit set of an interior trajectory is an equilibrium, a periodic orbit, or a closed chain of equilibria joined by connecting trajectories. Periodic orbits are excluded. Any periodic orbit of a planar system encloses an equilibrium, and for $H\le0$, there is no interior equilibrium for such an orbit to enclose. For $H>0$, the Lotka--Volterra equivalence used for the exploitation regime allows periodic orbits only when the interior equilibrium is a center with $\operatorname{tr}J(x^\ast)=0$~\citep{Hofbauer1981,Bomze1983}, which \eqref{eq:trace2} excludes.

No interior trajectory converges to a vertex. Both defenders and the public can invade the all-disruptor state, so it repels in every direction within the simplex. Only trajectories on the $D$--$A$ edge can approach the all-defender state, and only trajectories on the $D$--$P$ edge can approach the all-public state.

These edge directions also rule out a closed chain through the vertices. The only connecting trajectories that enter a vertex are the edges $A\to D$ and $D\to P$. They cannot complete a cycle, because trajectories on both sides of the $P$--$A$ edge approach the coalition. A chain avoiding the vertices would have to leave and re-enter $x^\dagger$ or $x^\ast$ through the interior, and none can. The interior equilibrium, whenever it exists, is a sink, and a sink has no outgoing connecting trajectory. The coalition is a sink for $H<0$; at $H=0$, \eqref{eq:centermanifold} shows that nearby interior trajectories approach it, so none leaves it; and for $H>0$, the only trajectories converging to it lie on the edge.

The $\omega$-limit set is therefore a single equilibrium. For $H>0$, it cannot be $x^\dagger$, so it is $x^\ast$. For $H\le0$, it is $x^\dagger$, the only remaining candidate. We summarize:

\medskip
\noindent\textbf{Proposition 2 (Exclusion or coexistence).} \emph{Under $M^{(2)}$, let $H=b_Dc_A-c_P(c_A+a_D)$, so that $H<0$ is equivalent to $c_P>c_P^\ast=b_Dc_A/(c_A+a_D)$.
\begin{enumerate}
\item[(i)] If $H<0$, there is no interior equilibrium, the coalition $x^\dagger$ of \eqref{eq:coalition} is a hyperbolic sink, and every interior trajectory converges to it. Defenders are excluded and the disruptor share is $x_A^\dagger=c_A/(c_P+c_A)$.
\item[(ii)] If $H=0$, there is no interior equilibrium; $x^\dagger$ is not hyperbolic, but every interior trajectory still converges to it, at an algebraic rate.
\item[(iii)] If $H>0$, the unique interior equilibrium $x^\ast$ of \eqref{eq:xstar2} is a sink and every interior trajectory converges to it; the coalition is a saddle. All three behaviors coexist, with defender share $x_D^\ast=H/S$ and disruptor share $x_A^\ast=N_A/S$.
\end{enumerate}
As $c_P$ increases through $c_P^\ast$, so that $H$ decreases through zero, the interior equilibrium passes through the coalition and exchanges stability with it: a transcritical bifurcation on the boundary of the simplex.}

\subsection{The accommodation regime: defender persistence}

A larger disadvantage for defenders among non-punishers, $a_D$, lowers the accommodation advantage needed for exclusion, holding the other payoffs fixed. Exclusion requires $c_P\ge c_P^{\ast}=b_D/(1+a_D/c_A)$, so the public's relative advantage from accommodation need not exceed the defenders' head-to-head advantage over disruptors. In the plane of the two dimensionless ratios $a_D/c_A$ and $c_P/b_D$, the threshold $H=0$ is the curve $c_P/b_D=1/(1+a_D/c_A)$, which falls toward zero as $a_D/c_A$ increases (Figure~\ref{fig:accommodation}a). The band between that curve and the line $c_P=b_D$ is the gap between the exact threshold and the stronger sufficient condition.

Two further consequences follow from \eqref{eq:N}. When $H>0$, the coalition is a saddle rather than a competing attractor, and the stable three-strategy equilibrium at the same payoff parameters has a smaller disruptor share than the coalition,
\begin{equation}
x_A^\dagger-x_A^\ast=\frac{(a_A+b_P+c_A)\,H}{(c_A+c_P)\,S}>0.
\label{eq:disruptorshare}
\end{equation}
The two shares meet at the bifurcation. The defender share at coexistence, $x_D^\ast=H/S$, decreases in $c_P$ and in $a_D$ and increases in $b_D$, holding the other payoffs fixed. It approaches zero continuously at the threshold, so near $c_P^\ast$ defenders survive only as a small minority (Figure~\ref{fig:accommodation}b).

For populations that initially contain all three strategies, the outcome is one of two. At or above the threshold, defenders disappear and the public and disruptors coexist at the coalition. Below it, all three strategies coexist at the interior equilibrium (Figure~\ref{fig:accommodation}c,d).

\subsection{Strategy loss and resurgence in finite populations}
\label{sec:finite}
We now examine how strategy loss interrupts resurgence in a finite population and how often disruptor majorities recur when lost strategies can return. Proposition~1 permits a strategy to recover from an arbitrarily small population share. Along a trajectory approaching the cycle, successive minima of the disruptor share eventually fall below $1/N$ for any fixed population size $N$. In a population of $N$ actors, however, the smallest positive strategy share is $1/N$. Once the last actor using a strategy switches, that strategy disappears and cannot return through imitation alone. \secref{sec:methods} reviews the relevant results: with an unstable interior equilibrium, the time to the first loss typically grows only like $\log N$~\citep{Frey2010}, and the two stochastic May--Leonard versions reviewed there lose a species in finite time, whereas a version in which an extinct species can reappear sustains the cycling~\citep{BarendregtThomas2023}. We simulate imitation alone first and then add reintroduction, which allows a missing strategy to return. The simulations use random pairwise copying, in which an actor is more likely to copy a higher-payoff strategy but can also copy a lower-payoff one, at the exploitation-regime parameters of Figure~\ref{fig:regimes}a. Each run starts at integer counts approximating the interior equilibrium. \secref{sec:methods} gives the update probabilities and the conversion of simulated time to replicator units, in which we report all times.

The two experiments vary the probability $\mu$ of random strategy adoption at each update and ask different questions. Without reintroduction ($\mu=0$), the finite process eventually reaches a permanent single-strategy state. We ask when the first strategy disappears, which strategy is lost, which ultimately prevails, and how many resurgences occur before fixation. With reintroduction ($\mu>0$), we ask how often resurgences occur when the loss of a strategy need not be permanent. We count a disruptor resurgence whenever the disruptor share rises above one half after having fallen below one tenth. Figure~\ref{fig:finitepop} reports the results, first without reintroduction and then with it.

Without reintroduction ($\mu=0$), the resurgence cycle ends quickly. At $N=1000$, the first strategy is lost after a median of $\FPextMedianThousand$ replicator units. Increasing $N$ a hundredfold to $10^5$ raises that median only to $\FPextMedianHundredThousand$ units, consistent with the logarithmic growth expected when the interior equilibrium is unstable (Figure~\ref{fig:finitepop}a,d). These times are to the first loss of a strategy, not to the later fixation of a single strategy.

The public is the strategy most often lost first, in $\FPlostPThousand\%$ of runs at $N=1000$ and $\FPlostPHundredThousand\%$ at $N=10^5$. This first-loss pattern is consistent with the deterministic flow near the all-disruptor state. When rare there, the public strategy declines at a proportional rate approaching $a_P=3$, while the defender strategy, whose growth carries the population away from that state, grows at a proportional rate approaching only $b_D=0.5$. After the public is lost, defenders usually displace disruptors. The population reaches the all-defender state in $\FPfreezeDThousand\%$ of all runs at $N=1000$ and $\FPfreezeDHundredThousand\%$ at $N=10^5$. Counting resurgences as defined above, the population freezes after at most one resurgence in $\FPepAtMostOneThousand\%$ of runs at $N=1000$ and after a median of $\FPepMedianHundredThousand$ resurgences at $N=10^5$. At these two population sizes, the simulations typically sustain only one or two resurgences before fixation.

With reintroduction ($\mu>0$), disruptor resurgences are observed at every parameter combination studied (Figure~\ref{fig:finitepop}b,c,e). With $\mu>0$, the finite process is irreducible, meaning that any composition can eventually be reached from any other. Every composition is revisited with finite expected return time, guaranteeing eventual resurgence. Whether resurgence is frequent, or occurs within a given observation window, is the question the simulations address.

We summarize the recurrence by the pooled simulated time per resurgence. At each $(N,\mu)$, this is the total simulated time in the $\FPrepsB$ runs of \FPTB{} replicator units each, divided by the total number of resurgences observed in them, an inverse resurgence rate. It summarizes the events counted within the observation windows and is not the expected return time to any particular composition. At $N=1000$, the time per resurgence is $\FPperiodThousandEminusTwo$ replicator units at $\mu=10^{-2}$, $\FPperiodThousandEminusThree$ at $10^{-3}$, $\FPperiodThousandEminusFour$ at $10^{-4}$, and $\FPperiodThousandEminusFive$ at $10^{-5}$. The time per resurgence increases as $\mu$ decreases, but grows less than proportionally to $1/\mu$ over the sampled range. The estimate is sparse at the smallest population and rate, $N=\FPminEpisodesN$ and $\mu=\FPminEpisodesMu$, where the $\FPrepsB$ runs contain $\FPminEpisodes$ resurgences in total, including $\FPzeroEpisodeRuns$ runs with none (open symbol in Figure~\ref{fig:finitepop}e).

Across the simulated observation windows at $N\ge10^3$, disruptors hold a majority of the population for between $\FPfracHiMin\%$ and $\FPfracHiMax\%$ of the time, depending on $N$ and $\mu$. Both the time per resurgence and these fractions of time with a disruptor majority summarize the observation windows as simulated, initial transients included; we have not assessed whether the runs reach stationarity.

At the payoffs studied, the loss of strategies interrupts the indefinite resurgence permitted by the continuum model; across the population sizes examined, the median number of resurgences before fixation ranged from zero to two. Reintroduction prevents the permanent loss of a strategy and allows recurrence to continue. Reintroduction also remains part of the dynamics as the population grows, because each update is a random adoption with probability $\mu$ whatever the value of $N$. With $\mu$ held fixed as $N\to\infty$, the deterministic limit includes both imitation and reintroduction; it is a replicator--mutator equation rather than the replicator equation~\eqref{eq:replicator}, so the runs with reintroduction concern an extension of the original model in which lost strategies can return. We do not analyze this deterministic extension. The simulations quantify recurrent disruptor majorities without establishing whether the observed fluctuations reflect a deterministic periodic orbit, stochastic effects, or a combination of the two.

We studied one set of payoff parameters. Different payoffs could change which strategy is most likely to disappear first, which strategy ultimately takes over, and how many resurgences occur before fixation. In this finite imitation process without reintroduction, however, the population eventually reaches a permanent state in which everyone uses one strategy, whatever the payoffs. With reintroduction, a finite population returns to every composition whatever the payoffs, so the occurrence of resurgence by itself does not identify the heteroclinic mechanism.

\section{Discussion}
\label{sec:discussion}

Defenders' ability to defeat disruptors directly does not determine whether institutional defense prevails across the population. Defensive behavior also competes with a response that tolerates norm-breaking while penalizing those who resist it. In our model, where defenders bear costs that non-punishers avoid, that response can sustain disruption despite defenders' direct advantage over disruptors. Our contribution is to connect two separately documented political tendencies---tolerance of norm-breaking and disapproval of confrontation---to a classical evolutionary game. The model assumes that the two tendencies occur together in one response and follows what happens as actors switch among defense, disruption, and that response.

Disruption can then persist in two ways. Under one set of payoffs, disruption repeatedly recovers after becoming rare. Defenders drive disruptors back, the non-punishing response then spreads at the defenders' expense, and disruption returns once defenders are scarce. This is the exploitation regime, in which condition~\eqref{eq:product} makes the classical May--Leonard cycle attracting (Proposition~1). In the accommodation regime, the outcome depends on the non-punishing response's advantage when rare among disruptors. At or above a threshold in that advantage, defenders disappear while the other two responses coexist. Below it, all three coexist (Proposition~2).

The threshold shows how limited the protection offered by head-to-head superiority can be. It lies strictly below the defenders' advantage over disruptors, so defenders can disappear even when that advantage exceeds the advantage the non-punishing response gains from accommodation. In the deterministic model, defenders' advantage over disruptors does not yield a stable all-defender outcome when all three strategies are initially present, because the public strategy can invade defenders.

We now consider three ways in which the model relates to empirical evidence. We will assess what the existing evidence supports, explain what the two payoff regimes mean empirically, and describe how the model could be tested.

First, the existing evidence suggests that voters' tolerance of norm-breaking varies with partisanship and the violation involved. In candidate-choice experiments, the estimated cost of an undemocratic position falls to a few percentage points of vote share in the scenarios most like real elections, and voters punish the other side far more readily than their own~\citep{GrahamSvolik2020}. Support for norm-eroding policies also rises when the respondent's own party holds power~\citep{SimonovitsMcCoyLittvay2022}. Other experiments find that voters penalize several specific violations across party lines, with the sharpest partisan split over voter-identification laws~\citep{CareyEtAl2022}.

The model combines weak punishment of norm-breakers with penalties on those who confront them. The studies reviewed in \secref{sec:intro} support these tendencies separately, but do not establish that the same actors reliably display both. The confrontation evidence also concerns particular tactics, such as incivility, extreme protest, and violence. Applying those findings to the defense of democratic institutions is another assumption. An online experiment offers a related behavioral pattern. Participants chose whether to join an optional public goods game. They later specified punishment for each possible contribution level in a compulsory game. Those who had opted out chose stronger punishment of the highest contribution than those who had participated~\citep{RandNowak2011}. This analogy does not establish that the same combination occurs in political settings.

Second, identifying the payoff regime requires knowing whether the public strategy can spread when rare among disruptors. Suppose nearly everyone uses the disruptor strategy. In the deterministic model, a rare public strategy declines under exploitation, where $M_{PA}<0$, and spreads under accommodation, where $M_{PA}>0$. The entry $M_{PA}$ compares the public strategy's payoff against disruptors with the payoff disruptors earn against one another. It does not measure whether institutional erosion improves or worsens an actor's welfare.

This distinction cannot be read directly from survey responses, although those responses could help explore the political interpretations of the regimes. Respondents who disapprove of democratic erosion yet impose little electoral punishment would fit the exploitation interpretation. Approval of norm-eroding measures when one's own party holds power would fit the accommodation interpretation~\citep{SimonovitsMcCoyLittvay2022}. Neither pattern determines the sign of $M_{PA}$. Identifying the regime would require a model connecting the observations to relative payoffs.

Third, testing the model requires observations of how strategies spread. The quantities to estimate are invasion fitnesses, the limiting proportional growth rates of a rare strategy as a population approaches a state occupied entirely by another strategy. These rates are the off-diagonal entries of $M$. One possible approach would follow actors over time, recording their strategies and switches between them. It would estimate how fast a rare strategy grows or declines where another dominates. Such an analysis would need to distinguish imitation from entry, exit, and other sources of strategy adoption, while keeping political circumstances comparable across settings.

These estimates would allow tests of both the regime distinction and the assumed competitive relations. Estimates of $M_{PA}$ would distinguish accommodation, where it is positive, from exploitation, where it is negative. Evidence that defenders can invade non-punishers, $M_{DP}>0$, or that rare non-punishers decline among defenders, $M_{PD}<0$, would contradict the assumed public--defender relation. These assessments would need to account for sampling uncertainty and the model's assumptions. Testing the product condition~\eqref{eq:product} would require all six expansion and contraction rates on a common time scale. The studies reviewed in \secref{sec:intro} measure attitudes and choices; none estimates these invasion rates.

Tests could also examine the model's predictions for changes in strategy shares over time. Under condition~\eqref{eq:product}, trajectories approaching the exploitation regime's attracting cycle show increasingly large fluctuations in those shares. Accommodation predicts convergence to an equilibrium, sometimes through oscillations that gradually diminish. Convergence alone cannot identify accommodation, because exploitation also produces it under the reverse inequality. Testing these predictions with electoral histories would require specifying how changes in strategy shares produce election outcomes. An incumbent's loss and return to office alone cannot identify the regime.

Our results have limitations of four kinds, concerning the payoffs, the rule by which strategies spread, the representation of political actors and audiences, and the size of the population.

First, the results depend on assumptions about the interaction payoffs. We keep the interaction payoffs in $M$ fixed as the strategy shares change. Each strategy's expected payoff still depends on the current population, but the matrix does not change as political institutions weaken or recover. Institutional change could alter the costs and rewards of defense, disruption, and the non-punishing response. Allowing the interaction payoffs to change would let the model describe that feedback.

The payoff assumptions also require that a defender among non-punishers earn a lower payoff than the non-punishers earn among themselves, even when disruptors are absent. Standing costs of vigilance or reputational penalties on a confrontational posture could produce this disadvantage, as \secref{sec:methods} explains, whereas a penalty paid only when a defender confronts a disruptor could not. This assumed payoff asymmetry has not been tested directly, and both regimes depend on it, because the cycle of the exploitation regime requires that defenders decline when rare among non-punishers and the threshold of the accommodation regime contains $a_D$. An extension could make the cost that defenders bear among non-punishers depend on the presence of disruptors, which would make the payoffs nonlinear in the strategy shares and require a separate analysis.

Second, we use proportional imitation, in which actors copy a better-performing actor they encounter with a probability proportional to the payoff difference, as one possible explanation for how strategies spread. Under the resulting replicator dynamics, strategies with above-average payoffs become more common. The evidence reviewed in \secref{sec:methods} shows that political behavior can spread through social influence, but it does not establish this specific relationship between payoff and spread. Imitation does not require actors to identify the best strategy across the whole population; they may instead respond to the apparent success of the actors they encounter.

Third, the model pools actors with different powers and opportunities to act. The empirical literature on backsliding distinguishes elites who hold or seek state power, the societal actors who link them to citizens, and citizens themselves, each with its own repertoire of erosive behaviors~\citep{Druckman2024}. Our population shares record behavioral responses without distinguishing these groups. Audience members belong to this same population, including actors using the non-punishing strategy. Their reactions enter the interaction payoffs, but the model does not separately describe how those reactions change.

These simplifications suggest two possible extensions. One would model elites and citizens as separate populations, with elite strategies describing institutional defense or disruption and citizen strategies describing responses to norm-breaking and confrontation. Another would place the audience outside the population of competing strategies. Either extension could retain the assumed payoff relations. Whether the same long-run outcomes would follow remains an open question.

Fourth, finite populations permit the complete loss of a strategy. The deterministic model allows a strategy to recover from an arbitrarily small positive share, whereas in a finite population that strategy disappears when its last actor switches. At the cyclic payoffs examined in the finite-population simulations, imitation alone supports only limited resurgence before the population reaches a permanent single-strategy state; the public strategy is usually lost first, and defenders usually take over. Reintroduction allows lost strategies to return. With continuing reintroduction, this finite process eventually revisits every composition, so recurrence alone does not identify the heteroclinic mechanism. The simulations measure how frequently disruptor resurgences occur at the chosen payoffs and over the observation windows. In this finite model, indefinite resurgence requires a continuing source of strategies that have disappeared, such as new entrants, generational turnover, or experimentation.

Finite population size also matters for accommodation. An attracting interior equilibrium can support long-lived coexistence in a finite population. In the stochastic models reviewed in \secref{sec:methods}, the mean time to strategy loss grows exponentially with population size when the interior equilibrium attracts~\citep{Frey2010}. This suggests that three-strategy coexistence below the accommodation regime's exclusion threshold could be long-lived, although we have not established its persistence time. As the threshold is approached from the coexistence side, the equilibrium defender share becomes small and defenders may become vulnerable to loss. At or above the threshold, the deterministic model itself predicts the loss of defenders. Once defenders are absent, the finite imitation process without reintroduction would eventually lose either the public strategy or the disruptor strategy as well. The deterministic coalition would therefore describe coexistence before eventual fixation, rather than permanent coexistence in that finite process.

Within these assumptions, the mechanism does not require the actors using the public strategy to oppose democracy. Withholding punishment from norm-breakers while penalizing those who confront them can sustain disruption despite the defenders' direct advantage over disruptors. The model does not explain why actors adopt this response.

\section{Methods}
\label{sec:methods}

\subsection{Actors, strategies, and the motivation of the payoff assumptions}
\label{sec:modelmethods}

The model has a single population of political actors, each using one of three strategies at a time. The population pools the three actor types that a review of the backsliding literature distinguishes: elites who hold or seek state power, the societal actors who link them to citizens, and citizens themselves~\citep{Druckman2024}. Actors can switch among these strategies without changing their institutional roles. A citizen who has excused norm-breaking can begin to oppose it or to take part in it, and an official who has taken part in it can begin to resist it. The state variables track how common each response is, not who performs it, and \secref{sec:discussion} considers what this pooling leaves out.

The institutional defender, $D$, opposes norm-breaking and acts to preserve democratic institutions, sometimes through visible confrontation such as public criticism, protest, litigation, or oversight. The anti-institutional disruptor, $A$, takes part in norm-breaking that weakens democratic institutions, for partisan or personal advantage. Our main motivating case is authoritarian disruption that turns legal and institutional tools against democratic constraints~\citep{LevitskyWay2002,Scheppele2018,Varol2015,LandauDixon2020}. The non-punishing public, $P$, withholds sanction from norm-breakers and instead penalizes those who visibly confront them. \secref{sec:intro} reviews the evidence for each tendency and states the assumption that they occur together.

On the defender--disruptor edge, the assumption $D\succ A$ is deliberately favorable to the defenders. Democratic resilience has two stages, averting the onset of autocratization and avoiding democratic breakdown once it begins, and democracies succeed far more often at the first than at the second~\citep{BoeseEtAl2021}. Democratic breakdowns alone do not establish whether defenders lacked a head-to-head advantage over disruptors or failed despite having one. Our model examines the latter possibility.

On the public--defender edge, the model assumes $P\succ D$, with $M_{PD}=b_P=U_{PD}-U_{DD}>0$ and $M_{DP}=-a_D=U_{DP}-U_{PP}<0$. A standing cost of institutional defense, borne whether or not disruptors are present, can produce both signs. The time and resources that vigilance requires lower a defender's payoff in every pairing. Among defenders, an actor using $P$ avoids that cost and can therefore earn more than the defenders earn among themselves, giving $b_P=U_{PD}-U_{DD}>0$. Among non-punishers, a defender bears the cost that the non-punishers avoid, and if no benefit offsets it, then $U_{DP}<U_{PP}$, giving $a_D>0$. This is the sense in which the public resembles the second-order free rider described in \secref{sec:intro}, benefiting from whatever defense occurs without paying for it. A penalty on confrontation can add to $a_D$ if it attaches to the defender strategy itself, as a reputational cost of a confrontational posture, rather than only to episodes of confrontation with a disruptor. In either form, the model assumes that defenders bear a net cost among non-punishers even when there is nothing to confront. This assumption goes beyond the two documented tendencies reviewed in \secref{sec:intro}, and the results depend on it.

The studies of incivility and protest reviewed in \secref{sec:intro} motivate a source of such costs. They document penalties for particular confrontational tactics, especially incivility and violence~\citep{MutzReeves2005,FeinbergWiller2020,Wasow2020}, and conflict-averse citizens take part less in high-conflict forms of participation~\citep{UlbigFunk1999}. These studies do not estimate either normalized entry or test whether defenders bear a net cost among non-punishers when disruptors are absent. The model represents audience reactions through its payoffs. The audience consists of actors from the same population, including those using $P$, and its behavior is not tracked separately. \secref{sec:discussion} considers this simplification.

On the public--disruptor edge, disruptors can invade a population of non-punishers in both regimes. The political motivation is that a disruptor who breaks norms among actors who withhold sanction gains the partisan or personal advantage that norm-breaking offers, at little cost, and so does better than the non-punishers do among themselves.

In the exploitation regime, the political motivation is a public that bears the cost of erosion but punishes it weakly. Here the relevant welfare comparison is $U_{PA}<U_{PP}$, in which the public is worse off under disruption than in its own baseline state. This welfare comparison motivates the interpretation. The invasion condition of \secref{sec:results}, $U_{PA}<U_{AA}$, is a separate comparison. Voters punish anti-democratic conduct only weakly when partisan and policy stakes are high~\citep{GrahamSvolik2020}.

In the accommodation regime, the political motivation is a public that accommodates a disruptor in return for short-run security, access to power or resources, or policy benefits. Such accommodation can lend the disruptor legitimacy and lower the cost of norm-breaking. The evidence motivating this interpretation concerns tolerance of norm-breaking by one's own side. In the United States, the constraints that once checked hostile, sharply divided parties have weakened, creating incentives for political elites to stick with their side even when it challenges democratic norms~\citep{PiersonSchickler2020}. Partisanship loosens the standards that voters apply to their own side's representatives~\citep{IyengarEtAl2019}, and perceived threat can raise tolerance for illiberal measures~\citep{FeldmanStenner1997,HetheringtonSuhay2011,SimonovitsMcCoyLittvay2022}.

\subsection{Replicator dynamics and the spread of behavior}
\label{sec:replicator}

Replicator dynamics describe how strategies that perform well become more common in a population~\citep{TaylorJonker1978,HofbauerSigmund2003}. The framework comes from evolutionary biology~\citep{TaylorJonker1978}, but nothing in it requires biological reproduction, and it has been used widely to model social behavior, including the spread of cooperation and the enforcement of norms~\citep{BoydEtAl2003,HofbauerSigmund2003,Sandholm2010}. In this paper, ``evolutionary'' refers to the spread of behavior through learning and imitation. \secref{sec:results} defines the population shares, the simplex on which they move, and the replicator equation~\eqref{eq:replicator} that governs them. Here we verify the properties of that equation used there and explain its interpretation as a model of learning.

The replicator equation~\eqref{eq:replicator} keeps the state on the simplex. The shares continue to sum to one because
\begin{equation}
\sum_i \dot{x}_i = \sum_i x_i \pi_i(x) - \bar{\pi}(x)\sum_i x_i = \bar{\pi}(x)-\bar{\pi}(x) = 0.
\label{eq:conservation}
\end{equation}
When $x_i=0$, the replicator equation gives $\dot{x}_i=0$, so an initially absent strategy remains absent. A positive share stays positive at every finite time, although it may approach zero as time increases. The state therefore remains on the simplex, which gives the invariance of the vertices and edges stated in \secref{sec:results}.

The normalization of the payoff matrix in \secref{sec:results} rests on a second property~\citep{HofbauerSigmund2003}. Adding a constant $k_j$ to every entry in column $j$ changes each strategy's expected payoff and the population average by the same amount, $k_j x_j$, and these changes cancel in \eqref{eq:replicator}. Subtracting the diagonal entry of an unnormalized matrix $U$ from every entry in its column, as in \eqref{eq:U}, therefore leaves the dynamics unchanged and gives a matrix with zero diagonal.

Two models of learning produce the same equations but give them different interpretations. In one, an individual becomes more likely to repeat a strategy the better it has paid off. In a limit of frequent play with small adjustments, that individual's probabilities of choosing each strategy follow the equations~\citep{BorgersSarin1997}. In the other, individuals copy others whose strategies are doing better, and the equations then describe the shares of the population using each strategy~\citep{Schlag1998,HofbauerSigmund2003}. Because our state variables are population shares, we adopt the second interpretation, under which political behaviors spread through the imitation of successful strategies.

\emph{Proportional imitation}, a form of success-biased cultural transmission~\citep{BoydRicherson1985}, provides a behavioral explanation for the replicator equation. An actor samples another actor at random and, if the other's payoff is higher, copies that actor's strategy with probability proportional to the payoff difference~\citep{Schlag1998,HofbauerSigmund2003}. In the continuous-time formulation of this imitation rule, population shares follow the replicator equation, up to a constant rescaling of time. Experiments support the broader premise that behavior can spread socially. In an online experiment, a health behavior spread through social networks~\citep{Centola2010}, and field experiments on voting find that political behavior spreads socially too~\citep{BondEtAl2012,Nickerson2008}. None of these studies tests the proportional-imitation rule itself, which remains a modeling assumption.

Some of our results hold for any \emph{payoff-monotone} dynamics, in which equal payoffs give equal proportional growth rates, higher payoffs give higher rates, and absent strategies stay absent~\citep{Weibull1995,HofbauerSigmund2003}. On each edge of our model, the signs of the two normalized off-diagonal entries fix the qualitative flow under any such rule. Payoff-monotone dynamics also share the same interior equilibria, the points where all three strategies earn equal payoffs. They need not produce the same trajectories, however, or the same time spent near each vertex. Our long-run results are established for replicator dynamics and do not automatically extend to other payoff-monotone dynamics.

\subsection{Cyclic dominance and heteroclinic cycles}
\label{sec:cyclic}

When three strategies are arranged so that each beats one of the others and loses to the third, no strategy dominates, and the pairwise advantages form a cycle. This is the structure of the game rock--paper--scissors, and it is known as \emph{cyclic dominance}. Cyclic dominance organizes competition in many natural systems, including the mating strategies of side-blotched lizards, antibiotic competition among strains of bacteria, and the spatial competition of plants and marine organisms, as reviewed in~\citep{Szolnoki2014}. The exploitation regime of our model has this structure. The accommodation regime does not. There the public and disruptors can each \emph{invade} the other, that is, increase when rare in a population of the other, so the arrangement is no longer one of cyclic dominance.

In a three-strategy population governed by replicator dynamics, cyclic dominance can send the population on an endless circuit. The prototype, from ecology, is a model of three competing species~\citep{MayLeonard1975}, in which the population swings from almost all of one type, to almost all of the next, to almost all of the third, and back. In the case that concerns us, identified by the product condition \eqref{eq:product} in \secref{sec:results}, the same succession repeats, and each circuit brings the trajectory closer to the triangle's boundary, where it spends longer near each single-strategy equilibrium before moving on. The set the trajectory approaches is a \emph{heteroclinic cycle}, a closed chain of single-strategy equilibria joined by paths that run along the boundary from one to the next~\citep{Hofbauer1994}.

Whether this cycle attracts or repels nearby trajectories depends on the motion along the edges of the triangle~\citep{Hofbauer1994}. For three-strategy replicator games with linear payoffs from a fixed matrix and strict cyclic dominance, as in our exploitation regime, the expansion and contraction rates defined in \secref{sec:results} decide among three outcomes. If the product of the three contraction rates exceeds the product of the three expansion rates, the boundary cycle attracts~\citep{Hofbauer1994,Gaunersdorfer1992}. If the product of the expansion rates is larger, the cycle repels and the interior equilibrium attracts, so the population settles into a steady mixture of all three strategies. If the two products are equal, the interior equilibrium is surrounded by neutrally stable closed orbits, and away from the equilibrium the strategy shares oscillate without the fluctuations growing or shrinking~\citep{Zeeman1980,Bomze1983,Hofbauer1981}. The order in which the strategies replace one another is fixed by who beats whom. \hyperref[sec:exploitation]{The exploitation-regime subsection of Results} derives the attracting-cycle condition~\eqref{eq:product} and explains why comparing the two products yields these three cases. In the loner game reviewed in \secref{sec:intro}, by contrast, both eigenvalues of the dynamics linearized at the all-loner state vanish~\citep{HauertEtAl2002b}, so that state has no expansion or contraction rate in this sense.

We focus on the attracting cycle, in which the swings in strategy shares become progressively more extreme and each period of dominance lasts longer. The replicator equations describe a population so large that the shares change smoothly, and they are deterministic, meaning that from a given composition the population always moves the same way. Models of finite populations are usually stochastic, meaning that chance enters the dynamics. In the models we consider, one randomly chosen actor at a time may switch strategies, so that in a population of $N$ actors the shares change in random steps of $1/N$, and a strategy whose last actor switches is lost until something other than imitation restores it. May and Leonard themselves noted that if the variables count discrete individuals, one species takes over after a few cycles and the other two go extinct~\citep{MayLeonard1975}.

Later work has made the timing precise for particular stochastic cyclic models. In the well-mixed models studied in this literature, in which interaction partners are sampled at random from the whole population, the mean time until a strategy is lost from a population of size $N$ depends on which of the three outcomes occurs in the deterministic model. When the interior equilibrium repels, as it does when the heteroclinic cycle attracts, the time typically grows only like $\log N$, so a larger population does not prolong coexistence by much~\citep{Frey2010}. The time grows like $N$ when the equilibrium is surrounded by neutral closed orbits, the case analyzed in~\citep{ReichenbachMobiliaFrey2006}, and exponentially in $N$ when it attracts~\citep{Frey2010}.

Finite-population outcomes also depend on the update rule, the interaction structure, and whether lost strategies can return. In some cyclic games, average drift over population compositions favors the boundary in small populations even when the deterministic interior equilibrium attracts. It favors the interior only above a critical population size that depends on the payoffs and the update rule~\citep{ClaussenTraulsen2008}. In a spatial model with interactions between neighbors on a grid, slow movement permits coexistence in spiral patterns, whereas sufficiently rapid movement destroys coexistence~\citep{ReichenbachMobiliaFrey2007}. For the May--Leonard model itself, two stochastic versions provably lose at least one species in finite time. A version in which an extinct species can reappear, as by immigration, instead sustains the cycling, with a finite average time per cycle~\citep{BarendregtThomas2023}.

\subsection{Finite-population simulations}
\label{sec:simmethod}

We simulate a mean-field-payoff version of the local update process~\citep{TraulsenClaussenHauert2005} at the exploitation-regime parameters of Figure~\ref{fig:regimes}a. In each elementary step, a focal actor and a role model are drawn at random from the $N$ actors, and the focal actor adopts the role model's strategy with probability $\tfrac12+\tfrac12\,w\,(\pi_{\mathrm{model}}-\pi_{\mathrm{focal}})/\Delta\pi_{\max}$. Here $\pi_{\mathrm{model}}$ and $\pi_{\mathrm{focal}}$ are the mean-field payoffs $\pi_i(x)=(Mx)_i$ of the two strategies at the current composition $x$. The constant $\Delta\pi_{\max}$ is the largest payoff difference that can occur. The parameter $w$ sets the sensitivity of adoption to payoff differences, and we take $w=1$. Unlike proportional imitation, this rule lets an actor copy a lower-payoff strategy, but its large-population limit without reintroduction is the same replicator equation, as we state below. With probability $\mu$, the focal actor instead adopts a uniformly random strategy. This reintroduction rate stands for the entry of new actors, generational turnover, or experimentation. It acts whether or not every strategy is present.

One generation consists of $N$ elementary steps. Without reintroduction, the large-population limit follows the replicator equation~\eqref{eq:replicator} after multiplying time in generations by $w/\Delta\pi_{\max}$~\citep{TraulsenClaussenHauert2005}. For the parameters used here, $w=1$, and the largest possible payoff difference occurs at the all-disruptor state, where defenders earn $b_D=0.5$ and the public strategy earns $-a_P=-3$, so $\Delta\pi_{\max}=b_D+a_P=3.5$. One generation therefore corresponds to approximately $\FPscale$ replicator time units. We use this common time scale to compare the simulations with the deterministic trajectory. Every run starts at integer counts approximating the interior equilibrium. The simulation uses an exact event-driven algorithm, which skips the elementary steps that leave the counts unchanged and samples the waiting time to the next change from a geometric distribution.

\section{Data availability}
No empirical data were collected for this study. The simulation outputs underlying Figure~\ref{fig:finitepop} and the reported numerical results, as per-run and summary tables, are archived with the code in Zenodo at \url{https://doi.org/10.5281/zenodo.22735002}.

\section{Code availability}
The R scripts used for the numerical analyses and for Figures~2--4, together with the C++ simulator they generate and their reference outputs, are archived in Zenodo at \url{https://doi.org/10.5281/zenodo.22735002}~\citep{TopazCode2026} under the MIT License, without restriction. Figure~1 is drawn by the TikZ code in the manuscript source. The results were produced with R 4.5.3, using the packages Rcpp 1.1.0, future 1.70.0, and furrr 0.4.0 for the finite-population simulations. All parameter values and random seeds are fixed in the scripts, and the simulation parameters are reported in the manuscript.

\section{Acknowledgements}
The author used Claude (Anthropic) to assist with copyediting, debugging code, and checking for adherence to the journal's style and formatting guidelines. All analysis, claims, and conclusions are the author's own, and he takes full responsibility for them.

\section{Funding}
This research received no funding.

\section{Competing interests}
The author declares no competing financial or non-financial interests.

\section{Author contributions}
C.M.T. conceived the study, performed the analysis, and wrote the manuscript.

\bibliographystyle{naturemag}
\bibliography{opinions}

\clearpage

\begin{figure}[p]
\centering
\begin{minipage}{0.48\linewidth}\centering
\resizebox{\linewidth}{!}{%
\begin{tikzpicture}[>=Stealth]
  \def\R{2.4cm}
  \node[draw, circle, thick, minimum size=11mm] (D) at (90:\R) {$D$};
  \node[draw, circle, thick, minimum size=11mm] (P) at (210:\R) {$P$};
  \node[draw, circle, thick, minimum size=11mm] (A) at (330:\R) {$A$};
  \node[font=\scriptsize\itshape,above=0.3mm of D] {defenders};
  \node[font=\scriptsize\itshape,below=0.3mm of P,align=center] {non-punishing\\public};
  \node[font=\scriptsize\itshape,below=0.3mm of A] {disruptors};
  \draw[->, thick] (D) -- node[pos=0.5, left=1mm, font=\scriptsize] {$P\succ D$} (P);
  \draw[->, thick] (P) -- node[pos=0.5, below=2.5mm, font=\scriptsize] {$A\succ P$} (A);
  \draw[->, thick] (A) -- node[pos=0.5, right=1mm, font=\scriptsize] {$D\succ A$} (D);
\end{tikzpicture}}\\[1mm]
{\footnotesize (a)}
\end{minipage}\hfill
\begin{minipage}{0.48\linewidth}\centering
\resizebox{\linewidth}{!}{%
\begin{tikzpicture}[>=Stealth]
  \def\R{2.4cm}
  \node[draw, circle, thick, minimum size=11mm] (D) at (90:\R) {$D$};
  \node[draw, circle, thick, minimum size=11mm] (P) at (210:\R) {$P$};
  \node[draw, circle, thick, minimum size=11mm] (A) at (330:\R) {$A$};
  \node[font=\scriptsize\itshape,above=0.3mm of D] {defenders};
  \node[font=\scriptsize\itshape,below=0.3mm of P,align=center] {non-punishing\\public};
  \node[font=\scriptsize\itshape,below=0.3mm of A] {disruptors};
  \draw[->, thick] (D) -- node[pos=0.5, left=1mm, font=\scriptsize] {$P\succ D$} (P);
  \draw[<->, thick, dashed] (P) -- node[pos=0.5, below=2.5mm, align=center, font=\scriptsize] {mutually\\invadable} (A);
  \draw[->, thick] (A) -- node[pos=0.5, right=1mm, font=\scriptsize] {$D\succ A$} (D);
\end{tikzpicture}}\\[1mm]
{\footnotesize (b)}
\end{minipage}
\caption{Schematic of the invasion relations in the two regimes. Vertices are the three strategies. A single-headed arrow points to the strategy that prevails on that edge, with $X\succ Y$ meaning that $X$ can invade $Y$ and $Y$ cannot invade $X$. The dashed double-headed arrow in panel~(b) denotes mutual invasibility, in which each strategy can invade the other and their mixed equilibrium is stable along that edge. (a)~In the \emph{exploitation} regime, the three relations form strict cyclic dominance, $D\to P\to A\to D$. (b)~In the \emph{accommodation} regime, only the $P$--$A$ edge changes.}
\label{fig:schematic}
\end{figure}

\clearpage
\begin{figure}[p]
\centering
\includegraphics[width=\linewidth]{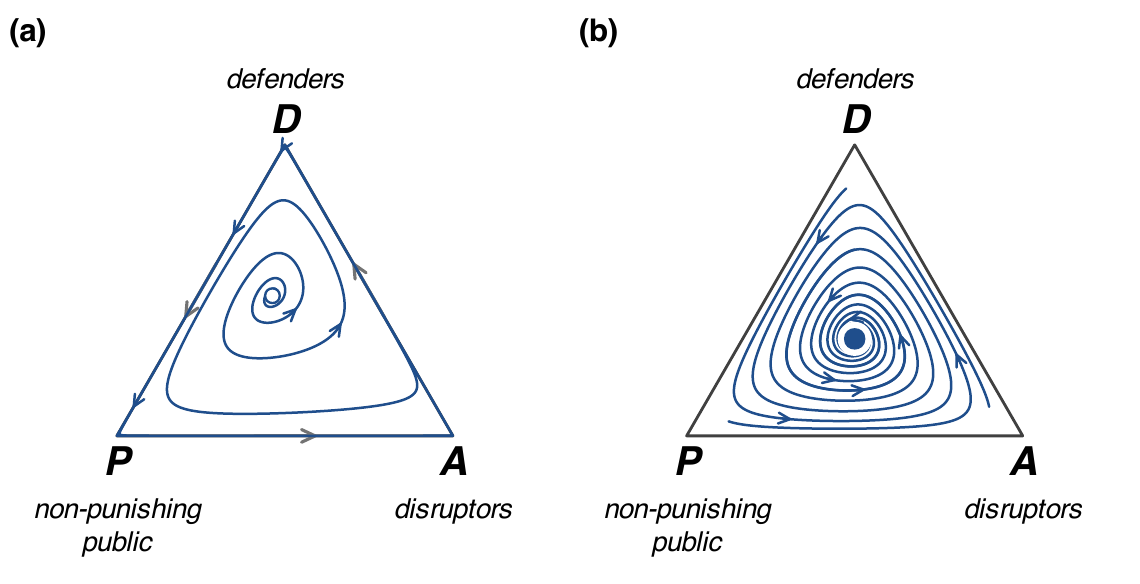}
\caption{Phase portraits of the exploitation regime, payoff matrix~\eqref{eq:M1}, on the simplex. The interior equilibrium is drawn as an open circle where it repels and as a filled circle where it attracts. \textbf{(a)}~Attracting heteroclinic cycle, $a_Pa_Aa_D=3>b_Pb_Ab_D=0.5$. The interior equilibrium repels, and a trajectory that starts near it spirals outward toward the boundary cycle. The gray edge arrows show the direction $D\to P\to A\to D$. The residence times near the vertices, which a phase portrait does not show, lengthen as the trajectory approaches the boundary. Parameters $a_D=1$, $b_D=0.5$, $a_P=3$, $b_P=1$, $a_A=1$, $b_A=1$. \textbf{(b)}~Attracting interior equilibrium, $a_Pa_Aa_D=1<b_Pb_Ab_D=3.375$. Trajectories from four initial conditions spiral into $x^\ast$, and the three behaviors coexist at fixed shares. Parameters $a_D=1$, $b_D=1.5$, $a_P=1$, $b_P=1.5$, $a_A=1$, $b_A=1.5$.}
\label{fig:regimes}
\end{figure}

\clearpage
\begin{figure}[p]
\centering
\includegraphics[width=0.88\linewidth]{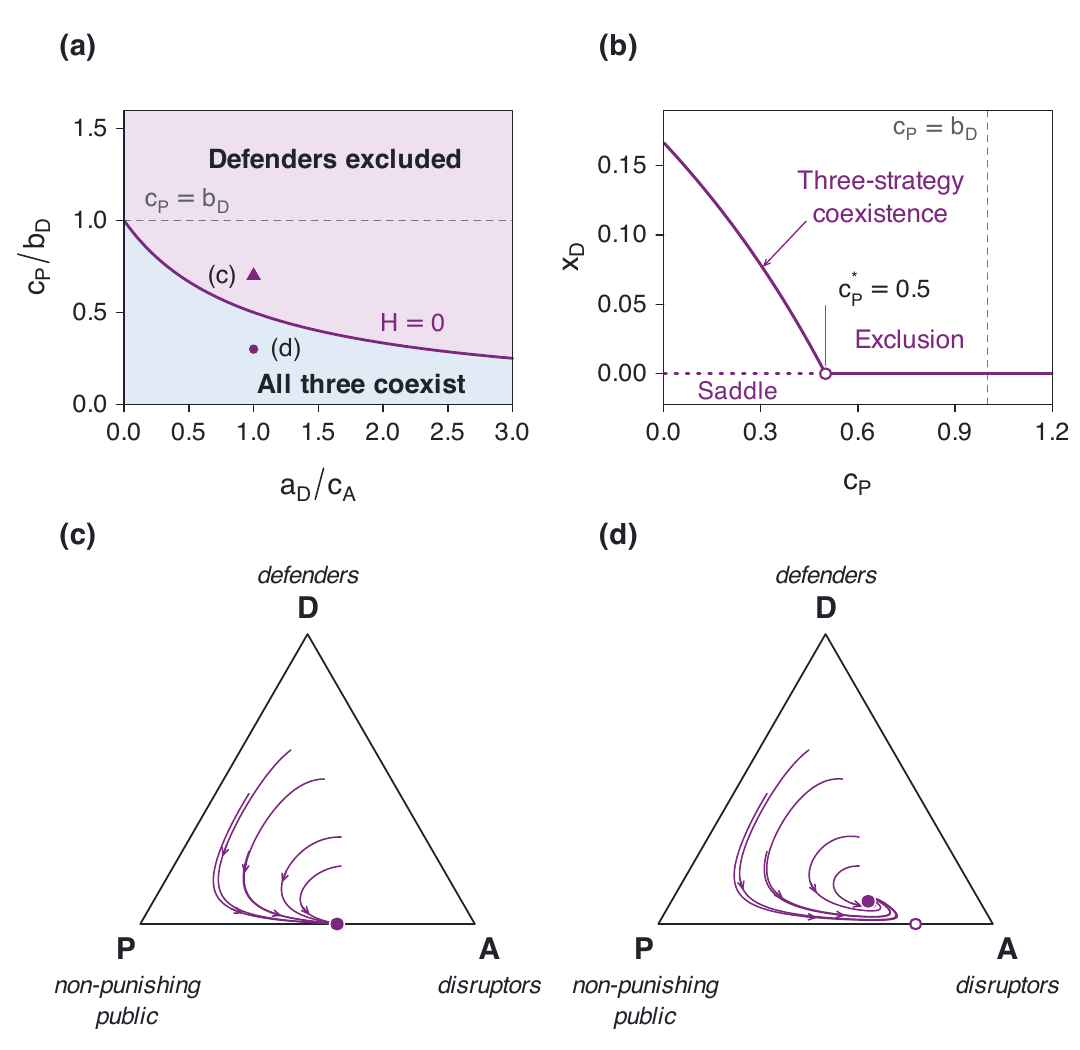}
\caption{The accommodation regime, payoff matrix~\eqref{eq:M2}. \textbf{(a)}~Regime diagram in the plane of the dimensionless ratios $a_D/c_A$ and $c_P/b_D$. On or above the curve $H=0$, that is, $c_P/b_D=1/(1+a_D/c_A)$, the coalition attracts every interior trajectory and defenders are excluded; below it, a unique interior equilibrium attracts every interior trajectory and the three behaviors coexist. The dashed line $c_P=b_D$ marks the boundary of the stronger sufficient condition $c_P>b_D$, and the band between the curve and the line is the additional exclusion region that the exact threshold establishes. The triangle and circle mark the parameters of panels (c) and (d). Panels (b)--(d) fix $a_D=b_D=b_P=a_A=c_A=1$, so that $c_P^\ast=0.5$. \textbf{(b)}~Bifurcation diagram, the equilibrium defender share as $c_P$ varies. The interior branch $x_D^\ast=H/S$ (solid) is attracting wherever it exists; its defender share reaches zero as $c_P$ increases to $c_P^\ast$, where the branch leaves the simplex through the coalition. The coalition branch $x_D=0$ is a saddle (dotted) for $c_P<c_P^\ast$ and attracting (solid) for $c_P>c_P^\ast$. The open circle marks $c_P^\ast$ itself, where the coalition is not hyperbolic but still attracts every interior trajectory, at an algebraic rate. The two branches exchange stability in a transcritical bifurcation. \textbf{(c)}~Phase portrait inside the band, $c_P=0.7$ ($H=-0.4$, $c_P<b_D$), in which every interior trajectory converges to the coalition $x^\dagger=(0,\,0.41,\,0.59)$. \textbf{(d)}~Phase portrait with $H>0$, $c_P=0.3$ ($H=0.4$), in which every interior trajectory converges to the interior equilibrium $x^\ast=(0.08,\,0.33,\,0.59)$ and the coalition (open circle) is a saddle. Filled circles mark attracting equilibria.}
\label{fig:accommodation}
\end{figure}

\clearpage
\begin{figure}[p]
\centering
\includegraphics[width=\linewidth]{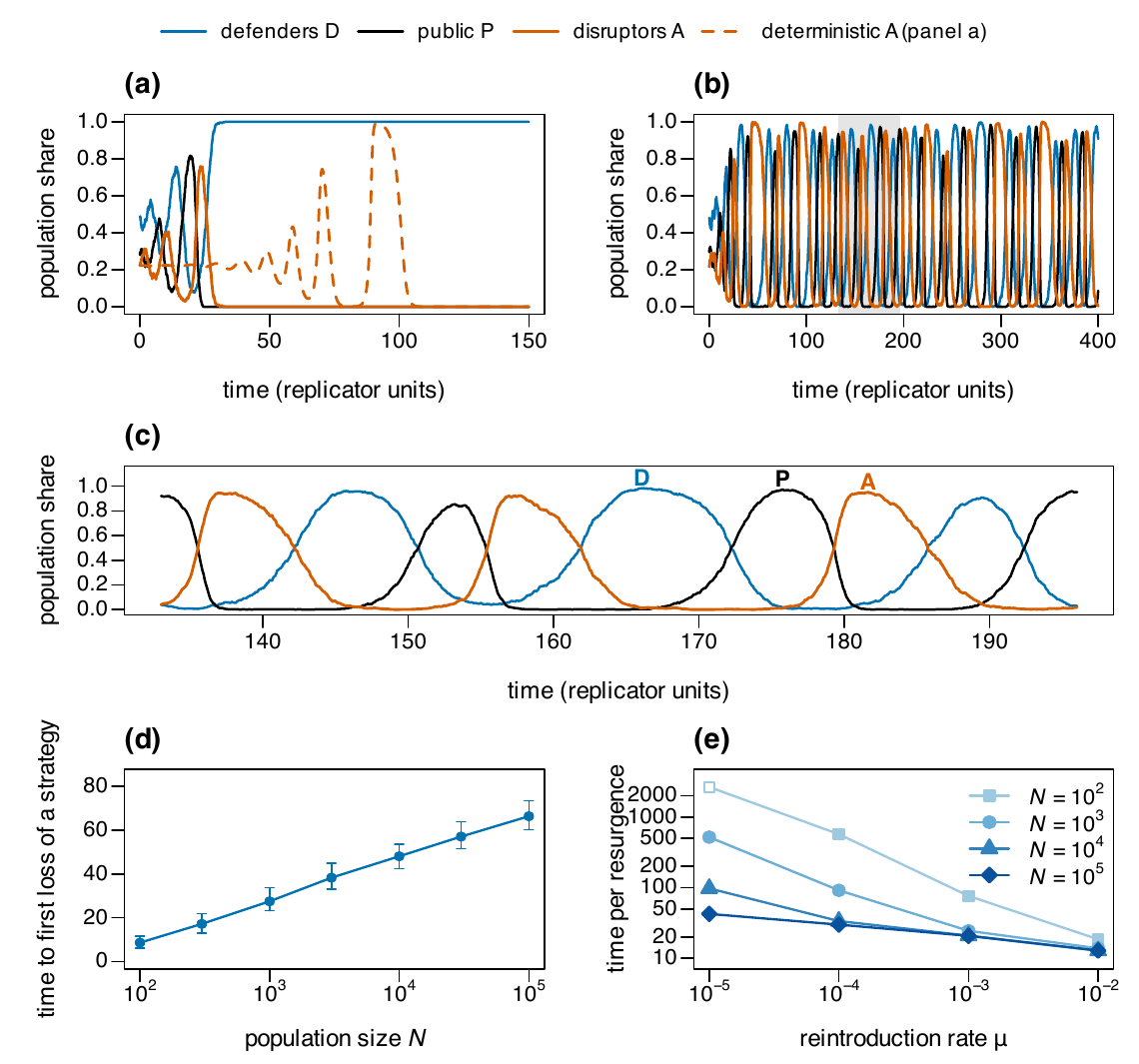}
\caption{Strategy loss and resurgence in finite populations, local update process at the exploitation-regime parameters of Figure~\ref{fig:regimes}a, time in replicator units. \textbf{(a)}~One run at $N=1000$ without reintroduction ($\mu=0$), selected to illustrate the most common outcome at this size. The public is lost at $t\approx\FPpanelALoss$, defenders then eliminate disruptors, and from $t\approx\FPpanelAFreeze$ the population is frozen at the all-defender state. The dashed curve is the deterministic disruptor share from the same initial condition. \textbf{(b)}~One run at $N=1000$ with reintroduction rate $\mu=10^{-3}$: disruptor resurgences continue throughout the displayed run, and the shaded interval is enlarged in panel (c). \textbf{(c)}~The shaded interval of panel (b), three successive rounds in which defenders, the public, and disruptors predominate in turn; the letters label the curves. \textbf{(d)}~Median time to the first loss of a strategy against $N$ at $\mu=0$, with interquartile range over $\FPrepsA$ runs. The growth with $N$ is consistent with $\log N$. \textbf{(e)}~Pooled simulated time per resurgence, an inverse resurgence rate, against $\mu$ for $N$ from $\FPNBmin$ to $\FPNBmax$, over $\FPrepsB$ runs of \FPTB{} replicator units each; open symbols mark cells with fewer than $\FPsparseThreshold$ resurgences in total. The time per resurgence lengthens as $\mu$ falls.}
\label{fig:finitepop}
\end{figure}

\end{document}